\documentclass[conference]{IEEEtran}
\IEEEoverridecommandlockouts
\usepackage{cite}
\usepackage{amsmath,amssymb,amsfonts}
\usepackage{algorithmic}
\usepackage{graphicx}
\usepackage{textcomp}
\usepackage{xcolor}
\usepackage{enumitem}

\def\BibTeX{{\rm B\kern-.05em{\sc i\kern-.025em b}\kern-.08em
    T\kern-.1667em\lower.7ex\hbox{E}\kern-.125emX}}

\begin{document}

\title{Trigger-DAQ and Slow Controls Systems in the Mu2e Experiment\\
}

\author{
\IEEEauthorblockN{1\textsuperscript{st} A.~Gioiosa}
\IEEEauthorblockA{\textit{Universit\`a di Pisa} \\
\textit{INFN Sezione di Pisa} \\
I-56127 Pisa, Italy \\
antonio.gioiosa@df.unipi.it}
\and
\IEEEauthorblockN{2\textsuperscript{nd} S.~Donati}
\IEEEauthorblockA{\textit{Universit\`a di Pisa} \\
\textit{INFN Sezione di Pisa} \\
I-56127 Pisa, Italy \\
}
\and
\IEEEauthorblockN{3\textsuperscript{rd} E.~Flumerfelt}
\IEEEauthorblockA{\textit{Fermi National Accelerator Laboratory} \\
Batavia, Illinois 60510, USA \\
}
\and
\IEEEauthorblockN{4\textsuperscript{th} G.~Horton-Smith}
\IEEEauthorblockA{\textit{Department of Physics} \\
\textit{Kansas State University}\\
Manhattan, Kansas 66506, USA \\
}
\and
\IEEEauthorblockN{5\textsuperscript{th} L.~Morescalchi}
\IEEEauthorblockA{\textit{INFN Sezione di Pisa} \\
I-56127 Pisa, Italy \\
}
\and
\IEEEauthorblockN{6\textsuperscript{th} V.~O'Dell}
\IEEEauthorblockA{\textit{Fermi National Accelerator Laboratory} \\
Batavia, Illinois 60510, USA \\
}
\and
\IEEEauthorblockN{7\textsuperscript{th} E.~Pedreschi}
\IEEEauthorblockA{\textit{INFN Sezione di Pisa} \\
I-56127 Pisa, Italy \\
}
\and
\IEEEauthorblockN{8\textsuperscript{th} G.~Pezzullo}
\IEEEauthorblockA{\textit{Yale University} \\
New Haven, Connecticut, 06520, USA \\
}
\and
\IEEEauthorblockN{9\textsuperscript{th} F.~Spinella}
\IEEEauthorblockA{\textit{INFN Sezione di Pisa} \\
I-56127 Pisa, Italy \\
}
\and
\IEEEauthorblockN{10\textsuperscript{th} L.~Uplegger}
\IEEEauthorblockA{\textit{Fermi National Accelerator Laboratory} \\
Batavia, Illinois 60510, USA \\
}
\and
\IEEEauthorblockN{11\textsuperscript{th} R.~A.~Rivera}
\IEEEauthorblockA{\textit{Fermi National Accelerator Laboratory} \\
Batavia, Illinois 60510, USA \\
}
}

\maketitle

\begin{abstract}

    The muon campus program at Fermilab includes the Mu2e experiment that will search for a charged-lepton flavor violating processes where a negative muon converts into an electron in the field of an aluminum nucleus, improving by four orders of magnitude the search sensitivity reached so far.
    
    Mu2e’s Trigger and Data Acquisition System (TDAQ) uses {\it otsdaq} as its solution. Developed at Fermilab, {\it otsdaq} uses the {\it artdaq} DAQ framework and {\it art} analysis framework, under-the-hood, for event transfer, filtering, and processing. 
    {\it otsdaq} is an online DAQ software suite with a focus on flexibility and scalability, while providing a multi-user, web-based, interface accessible through the Chrome or Firefox web browser. 
    The detector Read Out Controller (ROC), from the tracker and calorimeter, stream out zero-suppressed data continuously to the Data Transfer Controller (DTC). Data is then read over the PCIe bus to a software filter algorithm that selects events which are finally combined with the data flux that comes from a Cosmic Ray Veto System (CRV).
    A Detector Control System (DCS) for monitoring, controlling, alarming, and archiving has been developed using the Experimental  Physics  and  Industrial  Control  System (EPICS)  Open  Source  Platform. The DCS System has also been itegrated into {\it otsdaq}. The installation of the TDAQ and the DCS systems in the Mu2e building is planned for 2021-2022, and a prototype has been built at Fermilab’s Feynman Computing Center. We report here on the developments and achievements of the integration of Mu2e’s DCS system into the online {\it otsdaq} software.
\end{abstract}


\section{Introduction}
\begin{figure}[ht!]
    \begin{center}
        \includegraphics[width=0.48\textwidth]{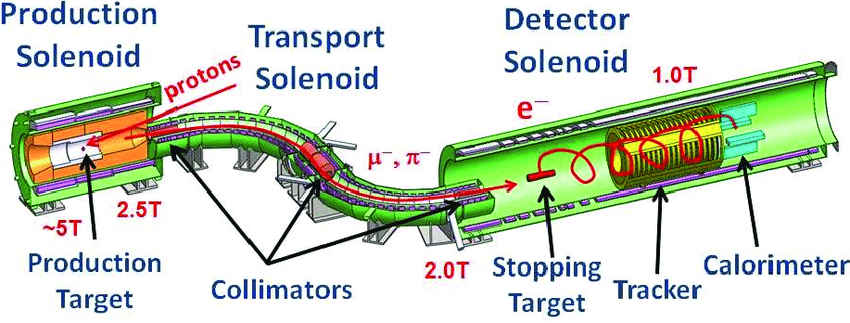}
        \caption{Mu2e experimental apparatus. The cosmic-ray veto system is not shown.}
   \label{fig:mu2e_layout}
   \end{center}
\end{figure}

Lepton flavor violation (LFV) has been observed in the neutral sector (neutrino oscillations), but not in the charged sector. In the Standard Model, the predicted rate of charged lepton flavor violating (CLFV) processes is below $10^{-50}$s~\cite{BOB}. However, many theories beyond the SM predict CLFV processes with rates observable by currently constructed HEP experiments~\cite{BOB}.
The Mu2e apparatus includes three superconducting solenoids (see Figure~\ref{fig:mu2e_layout}): 
(1) the production solenoid, where an 8 GeV proton pulsed-beam (period $\sim1.7$ $\rm\mu s$) hits a tungsten target, producing mostly pions; (2) the transport solenoid, which serves as a decay ``tunnel'' for the pions, and makes also charge and momentum selection, creating a low-momentum $\mu^-$ beam; (3) the detector solenoid, which houses an aluminum Stopping Target, where the muons get stopped and form muonic atoms, and the detector system (in a $1T$ solenoidal magnetic field) optimized to detect electrons and positrons from the muon conversions. The entire detector solenoid and half of the transport solenoid are covered with a cosmic-ray veto system (CRV), made out of 4-layers of extruded scintillator bars.

The detector consists of a 3.2 m long straw tube tracker and a crystal calorimeter. Both the tracker and the calorimeter have cylindrical symmetry. The inner part of each detector has an opening that provides for the free passage of non-interacting beam and makes the detectors ’’blind'' to low-momentum charged background particles produced in the stopping target.

The tracker consists of 36 equally spaced tracking planes, made out of 6 rotated panels arranged over two faces. A panel represents the basic unit of the tracker; it consists of 2 two staggered layers of straw tubes and has 96 straws in total. A more detailed description of the tracker detector can be found here~\cite{bartoszek2015mu2e,Pezzullo:2018rso}.

The crystal calorimeter is composed of two annuli with inner and outer radii of $37.4$ cm and $66$ cm respectively, filled by pure $CsI$ scintillating crystals and is placed downstream the tracker. Each annulus is composed of $674$ crystals of ($34 \times 34 \times 200$) $mm^3$ dimensions, each readout by two custom arrays of $2\times3$ 
$6 \times 6$ $mm^2$ UV-extended Silicon Photomultiplier (SiPM).
More details of the Mu2e calorimeter detector can be found in references:
\cite{Atanov2018}.
\section{The TDAQ System}
The Mu2e’s Trigger and Data Acquisition System (TDAQ) needs to satisfy the following requirements:
\begin{enumerate}
\item
  Provide efficiency better than 90\% for the signals;
\item
  Keep the trigger rate below a few kHz - equivalent to $\sim 7$ PB/year;
\item
  Achieve a processing time  $<5$ ms/event.
\end{enumerate}

\begin{figure}[ht!]
    \begin{center}
        \includegraphics[width=0.48\textwidth]{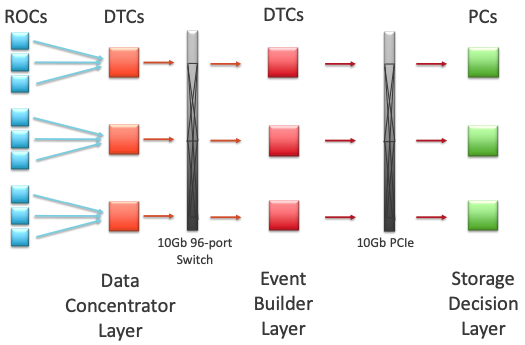}
        \caption{Mu2e data readout topology.}
   \label{fig:readout_topo}
   \end{center}
\end{figure}

The TDAQ uses {\it otsdaq} as a solution. Developed at Fermilab, it uses {\it artdaq}\cite{biery2018flexible, ARTDAQ} and {\it art}\cite{ART} software as event filtering and processing frameworks respectively. By incorporating {\it art} into the DAQ toolkit, complex trigger algorithms can be run on data in real-time, limited only by available processing power. 
{\it otsdaq} product allows rapid deployment, and it includes features such as a run control system using XDAQ~\cite{XDAQ} that allows artdaq-based DAQ systems to be used for development and calibration-mode runs at CMS. 
{\it otsdaq} has a library of supported front-end boards and firmware modules that implement a custom UDP protocol, and also support user defined front-end interface plugins. Additionally, an integrated Run Control Graphical User Interface (GUI) provides a multi-user, web-based controls and monitoring dashboard through the Chrome or Firefox web browser. In the case of Mu2e, a custom front-end interface plugin communicates with the detector Read Out Controller (ROC) firmware of the tracker and calorimeter through the Data Transfer Controller (DTC). ROCs stream out continuously the data, zero-suppressed, to the DTCs. The data of a given event is then grouped in a single server using a 10 GBytes switch. Then, the online reconstruction of the events starts and makes a trigger decision. If an event gets triggered, we pull also the data from the CRV and we aggregate them in a single data stream. Figure~\ref{fig:readout_topo} shows a scheme of the Mu2e data readout topology described above.
Data are saved in art’s ROOT format by default, reducing the amount of preprocessing necessary for offline analysis.
Current expected Mu2e data rate from front-ends is 40GBps and expected event size is 200KB.

The Mu2e main physics triggers use the info of the reconstructed tracks to make the final decision. The Mu2e Online track reconstruction is factorized into three main steps:
\begin{enumerate}[label=\roman*]
    \item Hits preparation, where the digitized signals from the sub-detectors are converted into reconstructed hits. 
    We use a Multi-Variate-Analysis (MVA) algorithm to identify and flag as “background” the hits compatible with being produced by a low-momentum (a few MeV) Compton $e^-$;
    \item Pattern-recognition to identify the group of hits that form helicoidal trajectories. 
    Two separate pattern-recognition methods are used: a calorimeter-seeded (calo-seeded) and a tracker-seeded (trk-seeded) algorithm;
    \item Track fit through the hit wires, which performs a more accurate reconstruction of the track. a simplified Kalman fit is performed to improve the accuracy in the reconstructed track parameters and thus - the background rejection. Resolution is about 1.4 MeV/c for the  conversion electron, which is about two times better than the momentum resolution after the pattern recognition.
\end{enumerate}
In the case of the conversion electron, the resulting efficiency is $\sim98 \%$ for the OR of the two algorithms, while it’s $\sim87\%$ or $\sim90\%$ if we consider only the calorimeter-seeded or the tracker-seeded algorithm respectively. 
The total instantaneous trigger rate is expected to be $\sim700Hz$. This number is dominated by low-momentum $e^-$ from the stopping target and that it can be reduced considerably either with a pre-scaler or by increasing the momentum threshold.
More details can be found here~\cite{Mu2eTracking}.
\section{The Distributed Control System}
\begin{figure}[ht!]
    \begin{center}
        \includegraphics[width=0.48\textwidth]{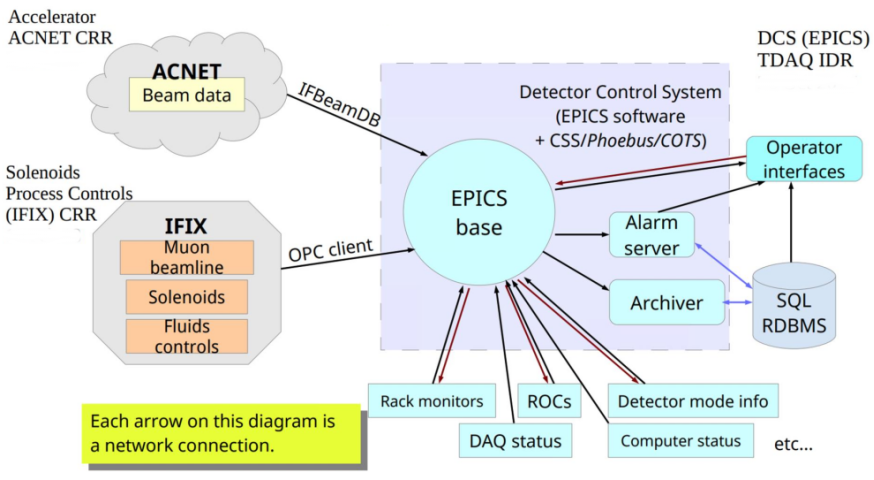}
        \caption{Mu2e DCS layout.}
   \label{fig:dcs_layout}
   \end{center}
\end{figure}

The Detector Control System (DCS) is the window, for experimenters and detector experts, on the status and health of the Mu2e detector. DCS must archive and present graphical user interfaces of both detailed and high-level displays of power supplies, liquid and gas system’s operational data, environmental temperatures and magnetic field strength, and status and run condition information for the data acquisition of every portion of the detector (see Figure~\ref{fig:dcs_layout}).

EPICS has been chosen for DCS implementation\cite{epics}. It is open source, originally developed at Argonne, and at Fermilab, has been heavily utilized in recent experiments\cite{epics_projects}. 
The system is designed to be robust enough to support a large number of
monitored variables and a broad range of monitoring and archiving rates and has to interface with a large number of systems to establish two-way communication for control and monitoring. 
The DCS supplies Ethernet networking and Ethernet-based controllers for general control and monitoring of power supplies, beamline equipment, and environmental sensors. It also provides generic analog and digital I/O endpoints. Other application-specific endpoints are provided by the corresponding subsystems. The DCS does not interface to safety-related controls or monitors for purposes other than secondary interlock (permit) and status logging.

The slow controls software includes the following components in order to provide complete monitoring and control of detector subsystems:
\begin{itemize}
    \item Control systems base that performs input and output operations and defines processing logic, scan conditions, alarm conditions, archiving rate, etc.;
    \item Alarm server that monitors all channels and sends alarm messages to operators;
    \item Data archiver that performs automatic sampling and storage of values for history tracking;
    \item Integrated operator interface that provides display panels for controls and monitoring.
\end{itemize}
The software has the ability to interface with external systems (e.g., cryogenics control system) and databases (e.g., beam database) to export data into slow controls process variables (or channels) for archiving and status displays. This allows integrating displays and warnings into one system for the experiment operators, and provides integrated archiving for sampled data in the archived database.

The archiver software allows data storage in an SQL database with adjustable rates and thresholds such that it is easy retrieve data for any channel by using channel name and time range. The alarm server software remembers state, support arbitrary number of clients, and provides logic for delayed alarms and acknowledging alarms. As part of the software, a standard naming convention for channels is followed to aid dealing with large number of channels and subsystems.

An input output controller (IOC), running for each subsystem on a central DAQ server, will provide soft channels for all data. 
Total channel count will be dominated by CRV, tracker, and calorimeter individual channels. We expect that the total number of slow control quantities will be $\sim30$ thousand. These quantities are expected to update on average no faster than twice per minute that will lead to a data rate in the range of tens of kilobytes per second.

As part of Mu2e DCS, {\it otsdaq} is responsible for delivering slow controls data from the DTCs and ROCs to EPICS. For example, each ROC may have a microcontroller, which handles DCS ’’slow control'' operations and is responsible for initializing the FPGA. The microcontroller is integrated into the FPGA. DCS commands can then be used to remotely load new software/firmware into the application program memory.

{\it otsdaq} allows the user to monitor or interact with their own DAQ hardware and all other devices managed by EPICS:
\begin{enumerate}
    \item Observe Process Variables (PVs) informations such as settings, alarms, warnings, readouts, timestamps, status;
    \item Interact through a web interface that is lightweight, user-Friendly, plug n’ play, customizable;
    \item Implement custom handling of PV alarms integrated with the TDAQ state machine transitions.
\end{enumerate}

A DCS web GUI has been developed, fully integrated into {\it otsdaq}. It includes a slow controls dashboard that has the following features: 
a searchable library of widgets, template widgets for customization, quick-snap pages for easy viewing, UI Scalability for high pixel screens or TVs, variable polling rates, Drag n’ Drop widgets, snap grid, infinite UI color customization, simple notes and names incorporated into the settings of each widget, page saving and loading.
It also includes configurable system message alarm notifications that use web and mail services to broadcast alerts to users.
\section{Conclusions}
We have presented the Trigger and Data Acquisition System (TDAQ) and Detector Control System (DCS) under development for the Mu2e experiment at Fermilab. 
TDAQ uses {\it otsdaq} an online DAQ software suite with a focus on flexibility and scalability, developed at Fermilab. 
DCS is based on EPICS, an open source framework developed at Argonne, and at Fermilab, that has been heavily utilized in recent experiments. 
{\it otsdaq} system includes a part of DCS that communicates with EPICS.

The modularity of {\it otsdaq} system allows easy integration into the core framework, so a Run Control Graphical User Interface (GUI) has been developed and integrated. It provides a multi-user, web-based controls and monitoring dashboard.
\section{Acknowledgements}
We are grateful for the vital contributions of the Fermilab staff and the technical staff of the participating institutions. This work was supported by the US Department of Energy; the Istituto Nazionale di Fisica Nucleare, Italy; the Science and Technology Facilities Council, UK; the Ministry of Education and Science, Russian Federation; the National Science Foundation, USA; the Thousand Talents Plan, China; the Helmholtz Association, Germany; and the EU Horizon 2020 Research and Innovation Program under the Marie Sklodowska-Curie Grant Agreement No. 690835, 734303, 822185, 858199. This document was prepared by members of the Mu2e Collaboration using the resources of the Fermi National Accelerator Laboratory (Fermilab), a U.S. Department of Energy, Office of Science, HEP User Facility. Fermilab is managed by Fermi Research Alliance, LLC (FRA), acting under Contract No. DE-AC02-07CH11359.

\end{document}